\documentclass{sf2a-conf}
\usepackage{graphicx}
\begin{document}
\TitreGlobal{SF2A 2008}
\title{Is it necessary to go beyond the ponctual mass approximation\\ for tidal perturber in close systems?}
\author{Mathis, S.}
\address{CEA/DSM/IRFU/SAp, CE Saclay, F-91191 Gif-sur-Yvette Cedex, France; AIM, UMR 7158, CEA - CNRS - Universit\'e Paris 7, France}
\author{Le Poncin-Lafitte, C.}
\address{Observatoire de Paris, SYRTE CNRS/UMR 8630, 61 Avenue de l'Observatoire, F-75014 Paris, France} 
\runningtitle{Is it necessary to go beyond the ponctual mass approximation for tidal perturber in close systems?}
\setcounter{page}{237} 
%
%
\maketitle
\begin{abstract}
With the discoveries of very close star-planet systems with planet orbiting sometimes at several star radius but also with well-known situations in our solar system where natural satellites are very close to their parent planet the validity of the ponctual mass approximation for the tidal perturber (respectively the parent star or planet when we study the close planet or natural satellite dynamics) has to be examined. In this short paper, we consider this problematic using results coming from a complete formalism that allows to treat the tidal interaction between extended bodies. We focus on its application to a simplified configuration.
\end{abstract}


\section{Context}
\par In celestial mechanics,  one of the main approximation done in the modelling of tidal effects (star-star, star-innermost planet or planet-natural satellites interactions) is to consider the tidal perturber as a point mass body. However a large number of extrasolar Jupiter-like planets orbiting their parent stars at a distance lower than 0.1 AU have been discovered during the past decade (Mayor {\it et al.}, 2005). Moreover, in Solar System, Phobos around Mars and the inner natural satellites of Jupiter, Saturn, Uranus and Neptune are very close to their parent planets. In such cases, the ratio of the perturber mean radius to the distance between the center of mass of the bodies can be not any more negligeable compared to 1. Furthermore, it can be also the case for very close but separated binary stars. In that situation, neglecting the extended character of the perturber have to be relaxed, so the tidal interaction between two extended bodies must be solved in a self-consistent way with taking into account the full gravitational potential of the extended perturber, generally expressed with some mass multipole moments, and then to consider their interaction with the tidally perturbed body. In the litterature, not so many studies have been done (Borderies, 1978-1980; Ilk, 1983; Borderies \& Yoder, 1990; Hartmann, Soffel \& Kioustelidis, 1994; Maciejewski, 1995).\\

\section{General formalism}
\par Several years ago, Hartmann, Soffel \& Kioustelidis (1994) introduced in Celestial Mechanics an interesting tool, based on Cartesian Symmetric Trace Free (STF) tensors, to treat straighforwardly the couplings between the gravitational fields of extended bodies. These tensors are fully equivalent to usual spherical harmonics but in addition a set of STF tensors represents an irreductible basis of the rotation group SO3 (Courant \& Hilbert 1953). It means that using algebraic properties of STF tensors, these objects become a powerful tool to determine the coupling between spherical harmonics in an elegant and compact way. However, as these tensors are not widely used in celestial mechanics, Mathis \& Le Poncin-Lafitte (2008) (hereafter MLP08) first recall their definition and fundamental properties and stress their relation with usual spherical harmonics. Then, the multipole expansion of gravitational-type fields is treated. First, the well-known external field of such body is derived using STF tensors; classical identities are provided. Next, the mutual gravitational interaction between two extended bodies and the associated tidal interaction are derived. We show how the use of STF tensors leads to an analytical and compact treatment of the coupling of their gravitational fields. We deduce the general expressions of tidal and mutual interaction potentials expanded in spherical harmonics. Using the classical Kaula's transform (Kaula, 1962), we express them as a function of the Keplerian orbital elements of the body considered as the tidal perturber. These results are then used to derive the external gravitational potential of such tidally perturbed extended body. Introducing a third body, its mutual interaction potential with the previous tidally perturbed extended body is defined that allows us to derive the disturbing function using the results obtained with STF tensors and the Kaula's transform. The dynamical equations ruling the evolution of this system are obtained.

\section{Case of an extended axisymmetric deformed perturber}

Here, our goal is to quantify the term(s) of the disturbing function due to the non-ponctual behaviour of the perturber B and to compare it to the one in the ponctual mass case. 
\vspace{-0.1cm}
\begin{figure}[ht]
\begin{center}
\resizebox{0.700\textwidth}{!}{
\includegraphics[scale=0.5]{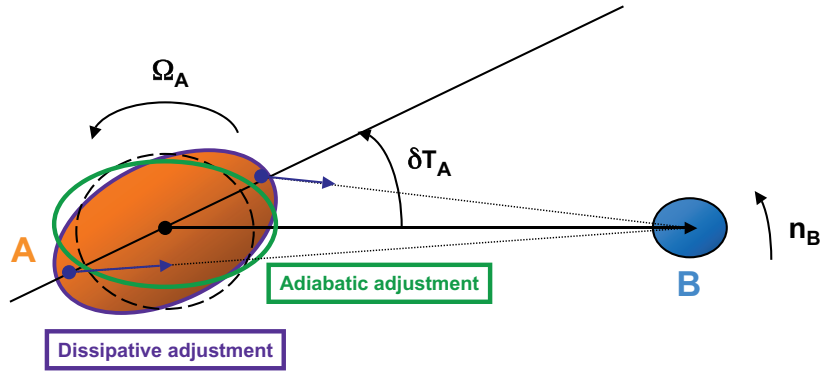}}
\caption{Classical tidal dynamical system. The extended body B is tidally disturbing the extended body A which adjusts itself with a phase lag $\delta_{{\rm T}_{\rm A}}$ due to its internal friction processes. The dynamics of B is then studied. $\Omega_{\rm A}$ and $n_{\rm B}$ are respectively the spin frequency of A and the mean motion of B.}
\label{fig1}
\end{center}
\end{figure}
\vspace{-0.5 cm}
To achieve this aim, some assumptions are done. First, we adopt the quadrupolar approximation for the response of A to the tidal excitation by B; thus, using the calculations of MLP08, the disturbing function is then given by
\begin{eqnarray}
{\mathcal R}&=&-\frac{G}{M_B}\frac{4\pi}{5}k_{2}^{\rm A}R_{\rm A}^5\sum_{m_{\rm A},l_{\rm B},m_{\rm B},j,p,q}\left\{\left\vert Z_{{\rm T}_{\rm A};2,m_{\rm A},L_{\rm I}}\left(\nu,K;\Psi_{2+l_{\rm B},m_{\rm A}+m_{\rm B},j,p,q}\right) \right\vert \left[\gamma_{l_{\rm B},m_{\rm B}}^{2,m_{\rm A}}\right]^2\left\vert M_{l_{\rm B},m_{\rm B}}^{\rm B} \right\vert ^2\right.\nonumber\\
& &\times\frac{1}{a_{\rm B}^{2\left(2+l_{\rm B}+1\right)}}\left[\kappa_{2+l_{\rm B},j}\right]^2\left[d_{j,m_{\rm A}+m_{\rm B}}^{2+l_{\rm B}}\left(\varepsilon_{\rm A}\right)\right]^{2}\left[F_{2+l_{\rm B},j,p}\left(I_{\rm B}\right)\right]^{2}\left[G_{2+l_{\rm B},p,q}\left(e_{\rm B}\right)\right]^{2}\nonumber\\
& &{\left.\times\exp\left[i\,\delta_{{\rm T}_{\rm A};2,m_{\rm A},L_{\rm I}}\left(\nu,K;\Psi_{2+l_{\rm B},m_{\rm A}+m_{\rm B},j,p,q}\right)\right]\right\}}
=\sum_{L_{\rm I}}{\mathcal R}_{L_{\rm I}},
\end{eqnarray}
where $L_{\rm I}=\left\{m_{\rm A},l_{\rm B},m_{\rm B},j,p,q\right\}$; indexes definition is given in MLP08.
$G$ is the universal constant of gravity. $M_{\rm B}$ is the mass of B and $R_{\rm A}$ is the radius of A. $k_{2}^{\rm A}$ is the Love's number of A that gives the linear adiabatic response of A to the tidal perturbation exerted by B. $Z_{{\rm T}_{\rm A}}$ describes the dissipation of the tidal kinetic energy by viscous friction and thermal diffusion ($\nu$ and $K$ are respectively the (turbulent) viscosity and the thermal diffusivity). On the other hand, the gravific potential of B is expanded as $V^{\rm B}\left(\bf r\right)=G\sum_{l_{\rm B}\ge0}M_{l_{\rm B},m_{\rm B}}^{\rm B}\,Y_{l_{\rm B},m_{\rm B}}\left(\theta,\varphi\right)/r^{l_{\rm B}+1}$ where the $M_{l_{\rm B},m_{\rm B}}^{\rm B}$ are the gravitational multipole moments of B and $\left(r,\theta,\varphi\right)$ are the spherical coordinates which have the center of mass of B for origin. $a_{\rm B}$, $e_{\rm B}$ and $I_{\rm B}$ are respectively the semi-major axis, the eccentricity and the inclination of the relative orbit of B. $\Psi_{L,M,j,p,q}$ is a function of the three other keplerian elements of this orbit, of the A angular velocity $\Omega_{\rm A}$ and of its precession angle. $\varepsilon_{\rm A}$ is the obliquity of A. $\gamma_{L_{1},M_{1}}^{L_{2},M_{2}}$ and $\kappa_{L,j}$ are coupling coefficients, which are detailed in MLP08. $d^{L}_{j,M}$, $F_{L,j,p}$ and $G_{L,p,q}$ are respectively obliquity, inclination and eccentricity special functions.\\

Since we are interested in the amplitude of ${\mathcal R}_{L_{\rm I}}$, we focus on its norm ($\vert{\mathcal R}_{L_{\rm I}}\vert$). On the other hand, as we know that the dissipative part of the tide is very small compared to the adiabatic one (cf. Zahn 1966), we can assume that $\left\vert Z_{{\rm T}_{\rm A}}\right\vert\approx1$ in this first step.

\begin{figure}[ht]
\begin{center}
\resizebox{0.65\textwidth}{!}{\includegraphics{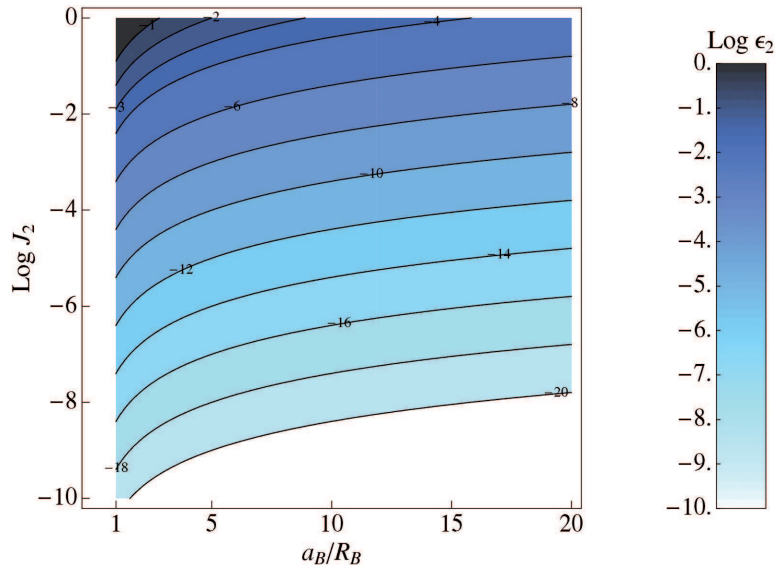}}
\caption{$\rm Log\,{\mathcal E}_{2}$ in function of $a_{\rm B}/R_{\rm B}$ and of $J_{2}$.}
\label{fig2}
\end{center}
\end{figure}

Let us first derive the terms $\vert{\mathcal R}_{L_{\rm I}}\vert$ due to the non-ponctual terms of the gravific potential of B which have a non-zero average in time over an orbital period of B, $\left<V^{\rm B}_{\rm N-P}\right>_{T_{\rm B}}\left(\bf r\right)=1/T_{\rm B}\int_{0}^{T_{\rm B}}V^{\rm B}_{\rm N-P}\left(t,\bf r\right){\rm d}t$ that corresponds to the axisymmetric rotational and permanent tidal deformations (see Zahn 1977) (the same procedure can of course be applied to the non-stationnary and non-axisymmetric deformations, but we choose here to focus only on $\left<V^{\rm B}_{\rm N-P}\right>_{T_{\rm B}}$ to illustrate our purpose). Then, as the considered deformations of B are axisymmetric, we can expand them using the usual gravitational moments of B ($J_{l_B}$) as
\begin{equation}
V^{\rm B}\left(\bf r\right)=\frac{GM_{\rm B}}{r}+\left<V^{\rm B}_{\rm N-P}\right>_{T_{\rm B}}\quad\hbox{where}\quad\left<V^{\rm B}_{\rm N-P}\right>_{T_{\rm B}}=G\sum_{l_{\rm B}>0}\left(M_{l_{\rm B},0}^{\rm S_{\rm B}}+M_{l_{\rm B},0}^{\rm T_{\rm B}}\right)\frac{Y_{l_{\rm B},0}\left(\theta,\varphi\right)}{r^{l_{\rm B}+1}},
\end{equation}
with
$
M_{l_{\rm B},0}^{\rm S_{\rm B}}+M_{l_{\rm B},0}^{\rm T_{\rm B}}=-\frac{J_{l_{\rm B}} M_{\rm B} R_{\rm B}^{l_{\rm B}}}{{\mathcal N}_{l_{\rm B}}^{0}}
$ where $M_{l_{\rm B},0}^{{\rm S}_{\rm B}}$ and $M_{l_{\rm B},0}^{{\rm T}_{\rm B}}$ are respectively the gravitational multipole moments induced by the permanent rotational and tidal deformations of B; $M_{\rm B}$ is the mass of B. ${\mathcal N}_{L,M}$ is the spherical harmonics ($Y_{L,M}$) normalization constant.
Then, we obtain
\begin{eqnarray}
\left\vert{\mathcal R}_{L_{\rm I}}^{J_{l_{\rm B}}}\left(a_{\rm B},e_{\rm B},I_{\rm B},\varepsilon_{\rm A}\right)\right\vert&=&\frac{G}{M_B}\frac{4\pi}{5}k_{2}^{\rm A}R_{\rm A}^5 \left[\gamma_{l_{\rm B},0}^{2,m_{\rm A}}\right]^2\left\vert M_{l_{\rm B},0}^{{\rm S}_{\rm B}} + M_{l_{\rm B},0}^{{\rm T}_{\rm B}} \right\vert ^2\nonumber\\
& &\times\frac{1}{a_{\rm B}^{2\left(2+l_{\rm B}+1\right)}}\left[\kappa_{2+l_{\rm B},j}\right]^2\left[d_{j,m_{\rm A}}^{2+l_{\rm B}}\left(\varepsilon_{\rm A}\right)\right]^{2}\left[F_{2+l_{\rm B},j,p}\left(I_{\rm B}\right)\right]^{2}\left[G_{2+l_{\rm B},p,q}\left(e_{\rm B}\right)\right]^{2}.
\label{App1}
\end{eqnarray}
On the other hand, the term $\vert{\mathcal R}_{L_{\rm I}}\vert$ associated to $M_{\rm B}$, namely the disturbing function in the case where B is assumed to be a ponctual mass, is given by:
\begin{eqnarray}
\left\vert{\mathcal R}_{L_{\rm I}}^{M_{\rm B}}\left(a_{\rm B},e_{\rm B},I_{\rm B},\varepsilon_{\rm A}\right)\right\vert&=&\frac{G}{M_B}\frac{4\pi}{5}k_{2}^{\rm A}R_{\rm A}^5 \left[\gamma_{0,0}^{2,m_{\rm A}}\right]^2\left\vert M_{0,0}^{{\rm S}_{\rm B}} \right\vert ^2\frac{1}{a_{\rm B}^{6}}\left[\kappa_{2,j}\right]^2\left[d_{j,m_{\rm A}}^{2}\left(\varepsilon_{\rm A}\right)\right]^{2}\left[F_{2,j,p}\left(I_{\rm B}\right)\right]^{2}\left[G_{2,p,q}\left(e_{\rm B}\right)\right]^{2}\nonumber\\
&=&\frac{G}{M_B}\frac{4\pi}{5}k_{2}^{\rm A}R_{\rm A}^5 M_{\rm B} \frac{1}{a_{\rm B}^{6}}\left[\kappa_{2,j}\right]^2\left[d_{j,m_{\rm A}}^{2}\left(\varepsilon_{\rm A}\right)\right]^{2}\left[F_{2,j,p}\left(I_{\rm B}\right)\right]^{2}\left[G_{2,p,q}\left(e_{\rm B}\right)\right]^{2}
\label{App2}
\end{eqnarray}
since $M_{0,0}^{\rm B}=\sqrt{4\pi}M_{\rm B}$. We now consider the ratio $\left\vert{\mathcal R}_{L_{\rm I}}^{J_{l_{\rm B}}}\right\vert / \left\vert{\mathcal R}_{L_{\rm I}}^{M_{\rm B}}\right\vert$, focusing on the configuration of minimum energy. In this state, the spins of A and B are aligned with the orbital one so that $\varepsilon_{\rm A}=I_{\rm B}=0$ (that leads to $j=m_{\rm A}$ and $p=\left(2-m_{\rm A}+l_{\rm B}\right)/2$) and the orbit is circular ($e_{\rm B}=0$). Then, we consider:
\begin{equation}
{\mathcal E}_{m_{\rm A},l_{\rm B}}=\frac{\left\vert{\mathcal R}_{L_{\rm I}}^{J_{l_{\rm B}}}\left(a_{\rm B},0,0,0\right)\right\vert}{\left\vert{\mathcal R}_{L_{\rm I}}^{M_{\rm B}}\left(a_{\rm B},0,0,0\right)\right\vert}.
\end{equation}
Using eqs. (\ref{App1}-\ref{App2}), we get its expression in function of $J_{l_{\rm B}}$ and of $\left(R_{\rm B}/a_{\rm B}\right)$: 
\begin{equation}
{\mathcal E}_{m_{\rm A},l_{\rm B}}=\frac{1}{4\pi}\left[\frac{1}{{\mathcal N}_{l_{\rm B}}^{0}}\frac{\gamma_{l_{\rm B},0}^{2,m_{\rm A}}}{\gamma_{0,0}^{2,m_{\rm A}}}\frac{\kappa_{2+l_{\rm B},m_{\rm A}}}{\kappa_{2,m_{\rm A}}}\frac{F_{2+l_{\rm B},2,\frac{l_{\rm B}}{2}}\left(0\right)}{F_{2,2,0}\left(0\right)}\right]^2J_{l_{\rm B}}^{2}\left(\frac{R_{\rm B}}{a_{\rm B}}\right)^{2 l_{\rm B}}\,.
\end{equation}
As it has been emphasized by Zahn (1966-1977), the main mode of the dissipative tide ruling the secular evolution of the system is $m_{\rm A}=2$. We thus define ${\mathcal E}_{l_{\rm B}}$ such that
\begin{equation}
{\mathcal E}_{l_{\rm B}}={\mathcal E}_{2,l_{\rm B}}=\left[\frac{1}{3}F_{2+l_{B},2,\frac{l_{B}}{2}}\left(0\right)\right]^{2}J_{l_{\rm B}}^{2}\left(\frac{R_{\rm B}}{a_{\rm B}}\right)^{2 l_{\rm B}}\,,
\end{equation}
which can be recast into
\begin{equation}
\log\left({\mathcal E}_{l_B}\right)=2\left[\log\left[\frac{1}{3}F_{2+l_{B},2,\frac{l_{B}}{2}}\left(0\right)\right]+\log J_{l_B}-l_{B}\log\left(\frac{a_B}{R_B}\right)\right]\,.
\end{equation}
Finally, keeping only into account the quadrupolar deformation of B ($J_{2}$), we get:
\begin{equation}
\log\left({\mathcal E}_{2}\right)=2\left[\log\left(\frac{5}{2}\right)+\log J_{2}-2\log\left(\frac{a_B}{R_B}\right)\right]\,.
\label{e2}
\end{equation}
This gives us the order of magnitude of the terms due to the non-ponctual behaviour of B compared to the one obtained in the ponctual mass approximation. It is directly proportional to the squarred $J_{2}$, thus increasing with $\varepsilon_{\Omega}^2$ (where $\varepsilon_{\Omega}={\Omega_{\rm B}^2}/{\Omega_{\rm c}^{2}}$, $\Omega_{\rm B}$ being the angular velocity of B and $\Omega_{\rm c}=\sqrt{\frac{G M_{\rm B}}{R_{\rm B}^{3}}}$) in the case of the rotation-induced deformation and with $\varepsilon_{\rm T}^{2}$ (where $\varepsilon_{\rm T}=q\left({R_{\rm B}}/{a_{\rm B}}\right)^3$ where $q=M_{\rm A}/M_{\rm B}$) in the tidal one, while it increases as $\left(R_{\rm B}/a_{\rm B}\right)^{4}$. Therefore, as it is shown in Fig. (\ref{fig2}), the non-ponctual terms have to be taken into account for strongly deformed perturbers ($J_{2}\ge10^{-2}$) in very close systems ($a_{\rm B}/R_{\rm B}\le5$) while they decrease rapidly otherwise. 

\vspace{-1.0 cm}
%
%

\end{document}